# Implementation of reflection matrix microscopy: An algorithm perspective

SUNGSAM KANG[1,2], SEOKCHAN YOON[3], AND WONSHIK CHOI[1,2,*]

[1]*Center for Molecular Spectroscopy and Dynamics, Institute for Basic Science, Seoul 02841, Korea*
[2]*Department of Physics, Korea University, Seoul 02855, Korea*
[3]*School of Biomedical Convergence Engineering, Pusan National University, Yangsan 50612, Korea*
*\*wonshik@korea.ac.kr*

**Abstract:** Over the past decade, reflection matrix microscopy (RMM) and advanced image reconstruction algorithms have emerged to address the fundamental imaging depth limitations of optical microscopy in thick biological tissues and complex media. In this study, we introduce significant advancements in reflection matrix processing algorithms, including logical indexing, power iterations, and low-frequency blocking. These enhance the processing speed of aperture synthesis, 3D image reconstruction, and aberration correction by orders of magnitude. Detailed algorithm implementations, along with experimental data, are provided to facilitate the widespread adoption of RMM in various deep-tissue imaging applications.

## 1. Introduction

Optical microscopy has been an indispensable tool in diverse areas of science and engineering. However, optical microscopy techniques experience a substantial reduction in resolution and imaging depth due to multiple light scattering and sample-induced aberrations in thick biological tissues or complex materials [1]. Light waves propagating through such media undergo random deflections of propagation angles and phase retardation depending on the propagation angle, resulting in distortions of the point-spread function and a decrease in peak intensity. To mitigate these issues, various adaptive optics (AO) methodologies have been employed [2]. AO techniques are broadly categorized into hardware-based and software-based approaches. Hardware-based methods utilize wavefront shaping devices to correct aberrations, which can be subdivided into wavefront sensing techniques [3-6] and sensorless techniques [7-9]. Wavefront sensing methods, such as those employing Shack-Hartmann wavefront sensors [10], require the presence of approximate point sources known as guide stars within the sample to measure aberrations. Sensorless techniques iteratively optimize an image metric through image acquisition, which can be time-consuming and necessitates that initial aberrations be mild enough for the object to be partially visible. Software-based approaches have mostly been applied to coherent imaging applications that utilize the interferometric detection of intrinsic reflectance [11-13]. This approach faces challenges in disentangling the aberrations of the incident and reflected waves since both have the same wavelength. One strategy to address this issue involves using a single-angle plane wave to eliminate the effect of incident aberrations, but it is prone to multiple scattering noise due to the absence of confocal gating.

To fundamentally overcome these limitations, an innovative approach known as reflection matrix microscopy (RMM) has been developed [14-19]. Previous AO methods are constrained mainly by capturing only partial information about light-sample interactions. In contrast, RMM captures the reflection signal for all detection channels across each incident channel. The representative realizations scan the incident angle or focus and measure the amplitude and phase of the reflected waves by an array detector over a wide field, providing much more comprehensive information than using single incident angles or confocal detection methods.

Arbitrary illumination patterns can also be used, such as speckle illuminations, as long as they can cover multiple orthogonal incident channels. Initially, time-gated reflection matrices were measured using broadband sources to attenuate multiple scattering with different flight times from those of signal waves [15]. Recent studies employed wavelength scanning to record spatio-spectral reflection matrices, allowing for recording volumetric responses [20].

Leveraging the reflection matrix, a pioneering algorithm called Closed-Loop Accumulation of Single Scattering (CLASS) has been developed to separate incident and reflected aberrations from the measured reflection matrix [16]. This algorithm exploits wave correlations based on the property that objects within a thin depth section, as defined by the objective focus and time-gating, exhibit spatial frequency shifts with respect to changes in the incident angle, eliminating the need for a guide star. The algorithm iteratively applies wave correlations among the incident and reflection channels, leading to the separate identification of input and output aberrations. Furthermore, the algorithm operates effectively even in the presence of strong multiple scattering noise by leveraging multiple incident angles, thus providing noise suppression capability equivalent to that of confocal microscopy. Consequently, this algorithm proves effective even in low-reflectance biological tissues. The applicability of this method to deep-tissue imaging has been demonstrated in vivo, in both mature zebrafish and through-skull deep-tissue imaging in mouse brains [17, 19]. Recent developments have extended this method to the frameworks of conjugate AO and multi-conjugate AO, enabling the correction of multiple scattering and converting them to signals [21, 22]. The ability to map out the sample-induced aberrations in biological tissues using intrinsic reflectance implies that this method can be used for wavefront sensing AO. Complex aberrations identified by reflection matrix microscopy were used to correct the excitation PSF in multiphoton imaging and the emission PSF in single-molecule localization microscopy to enhance their imaging depths [23]. The CLASS algorithm essentially performs matrix factorization, decomposing the input aberration matrix, object matrix, and output aberration matrix. This framework is so general that it has been extended to nonlinear coherent imaging, such as second harmonic generation (SHG) and third harmonic generation (THG) imaging, and incoherent imaging, such as multiphoton fluorescence imaging, to computationally correct sample-induced aberrations [24-28].

In this study, we report the algorithmic advancements designed to optimize the performance of the CLASS algorithm. To begin with, we provide a detailed overview of reflection matrix microscopy from an algorithmic perspective. On this basis, we propose advances in the core engine of the algorithm, which includes the efficient conversion of the reflection matrix to a shifted momentum basis by logical indexing and the wave correlation operations based on power iterations. These developments allow for an increased processing speed by orders of magnitude compared to the original implementations. Furthermore, we propose strategies to enhance the convergence rate and achieve three-dimensional image reconstruction. To facilitate the widespread adoption of the proposed algorithm to the deep-tissue imaging communities, we provide Matlab implementations, simulation data, and experimental results for validation [29].

The structure of this paper is as follows: Section 2 elucidates the principles of reflection matrix microscopy. Sections 3 and 4 detail the 2D and 3D image reconstruction processes in RMM. Section 5 introduces the CLASS algorithm and its implementation. Section 6 presents experimental validations and results. Additionally, Section 7 introduces the low-spatial frequency blocking method for better accuracy and faster convergence of the CLASS algorithm. Finally, Section 8 discusses the implications of our findings and potential future research directions in this domain.

## 2. Working principles of reflection matrix microscopy

A reflection matrix, denoted as $\mathbf{R}$, characterizes the impulse response of the electric field on the reflection side of a target sample under an orthogonal incident electric field basis, such as position or spatial frequency. As depicted in Fig. 1a, the matrix elements $r_{ji}$ represents the electric field reflected from the sample, measured at $j^{\text{th}}$ detection channel under $i^{\text{th}}$ illumination channel. Once the reflection matrix $\mathbf{R}$ is measured, the reflected field $\mathbf{E}_{\text{out}}$ from the sample under arbitrary incident field $\mathbf{E}_{\text{in}}$ can be reconstructed from $\mathbf{R}$ as $\mathbf{E}_{\text{out}} = \mathbf{R}\mathbf{E}_{\text{in}}$.

In practice, the reflection matrix requires measuring multiple complex electric field images under a set of orthogonal incident electric field bases whose basis number $N$ is set by the region of interest (ROI) of the sample and the numerical aperture of the objective lens. Since the reflection matrix requires to measure the electric field, interferometric detection is necessary [15]. Figure 1b-c describes the construction of the reflection matrix $\mathbf{R}$ by measuring the output electric field $\mathbf{E}_{\text{out}}$ for each input field $\mathbf{E}_{\text{in}}^{(i)}$.

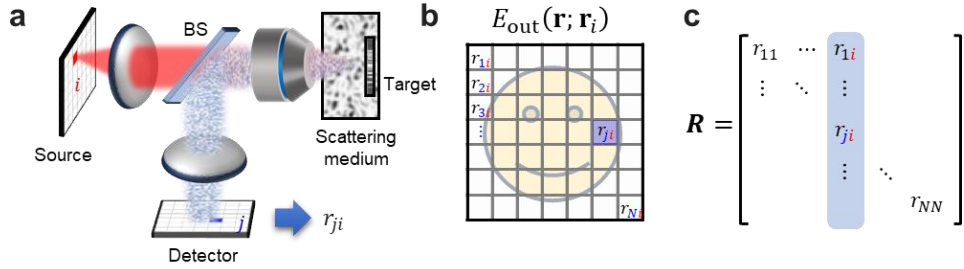

Figure 1. Schematic of reflection matrix microscopy. a, Concept of reflection matrix measurement. b-c, Mapping of reflection matrix by vectorizing detected 2D output electric field $\mathbf{E}_{\text{out}}(\mathbf{r}; \mathbf{r}_i)$ into a column vector.

## 3. Image reconstruction in reflection matrix microscopy

For an experimental measurement of the reflection matrix, we prepare the incident electric field basis, which can be a plane wave basis at different angles of incidence, a position basis by scanning the laser beam focus, or a set of predetermined random speckle patterns. In general, they can be expressed as the input basis $\{\mathcal{A}\} = \{\mathbf{E}_{\text{in},1}, \mathbf{E}_{\text{in},2}, \cdots, \mathbf{E}_{\text{in},i}, \cdots, \mathbf{E}_{\text{in},N}\}$. We then measure a set of corresponding reflected electric field $\{\mathcal{B}\} = \{\mathbf{E}_{\text{out},1}, \mathbf{E}_{\text{out},2}, \cdots, \mathbf{E}_{\text{out},i}, \cdots, \mathbf{E}_{\text{out},N}\}$ for a given ROI. In other words, $\mathcal{A}$ and $\mathcal{B}$ are a stack of complex field maps, as shown in Fig. 2a.

Let us consider the illumination and detection coordinates in the position domain. Then $\mathbf{E}_{\text{in},i}$ and $\mathbf{E}_{\text{out},i}$ are composed of the elements $E_{\text{in}}(x, y, i)$ and $E_{\text{out}}(x, y, i)$, respectively. Here, $x$ and $y$ are horizontal and vertical coordinates in the position domain. When the measured ROI is composed of $L \times M$ pixels, the dimensions of the image stack $\mathcal{A}$ and $\mathcal{B}$ are $L \times M \times N$. For computational purposes, we map the 3D image stacks into 2D matrices by vectorizing the individual complex images in the stack, as shown in Fig. 2b. In this way, $\mathcal{A}$ and $\mathcal{B}$ are converted to incident field matrix $\mathbf{A}$ and the reflected field image matrix $\mathbf{B}$. Then the dimensions of the two matrices $\mathbf{A}$ and $\mathbf{B}$ become $LM \times N$. In Matlab, this process can be done by simply reshaping $L \times M \times N$ array into $LM \times N$ array. Finally, the reflection matrix $\mathbf{R}$ whose dimension is $LM \times LM$ can be constructed by $\mathbf{R} = \mathbf{B} \times \mathbf{A}^{-1}$. If the camera is placed at the conjugate image plane of the target object, the reflection matrix $\boldsymbol{R}$ is defined in the position domain regardless of the input basis. As a result, the diagonal elements of $\boldsymbol{R}$ correspond to the confocal imaging of the target object with $L \times M$ as shown in the last image of Fig. 2b.

The optimum sample size $L$ and $M$ for a given ROI and spatial frequency bandwidth can be determined by Nyquist sampling theory. For instance, if the wavelength is $\lambda$ and the numerical aperture (NA) is $\alpha$, the maximum spatial frequency is $2\pi\alpha/\lambda$, and the Nyquist sampling rate will be $4\pi\alpha/\lambda$. If the physical dimension of ROI is $L\Delta x \times M\Delta y$, the Nyquist sampling theorem results in the horizontal and vertical sampling periods of $\Delta x = \Delta y = \lambda/2\alpha$. As a result, the confocal image reconstructed in Fig. 2b has a spatial resolution of $\lambda/2\alpha$. However, general confocal microscopy should have a better spatial resolution up to $\lambda/4\alpha$ due to the filtering process involving the illumination and detection pinhole. Specifically, the confocal filtering process can be described in the spatial frequency domain as a synthetic aperture process, which effectively doubles the spatial frequency bandwidth [30].

In reflection matrix microscopy, such high spatial frequency information can be obtained by reconstructing the object image in the spatial frequency domain. As described in Appendix A, the spatial frequency basis reflection matrix $\widetilde{\boldsymbol{R}}$ can be obtained by the Fourier transform of $\boldsymbol{R}$ as $\widetilde{\boldsymbol{R}} = \boldsymbol{DRD}^\dagger$ with a discrete Fourier transform matrix (DFM) $\boldsymbol{D}$. It is noteworthy that the use of the fast Fourier transform (FFT) algorithm is more efficient than the use of DFM. We provide the FFT algorithm-based Matlab codes for converting $\boldsymbol{R}$ into $\widetilde{\boldsymbol{R}}$, and its inverse process in the software package [29].

For a layer of the object $O(x, y)$ located at a focal depth $z$, the individual element of $\widetilde{\boldsymbol{R}}$ can be described by
$$\tilde{R}(\mathbf{k}_{\text{out}}; \mathbf{k}_{\text{in}}) = \gamma \tilde{O}(\mathbf{k}_{\text{out}} - \mathbf{k}_{\text{in}}) + \widetilde{M}(\mathbf{k}_{\text{out}}; \mathbf{k}_{\text{in}}). \tag{1}$$
Here, $\tilde{O}$ is the Fourier transform of the target object, $\gamma$ is the attenuation of the ballistic wave interacting with the target object, $\widetilde{M}$ is multiple scattering components, $\mathbf{k}_{\text{out}}$ and $\mathbf{k}_{\text{in}}$ are spatial frequency of reflected and incident field, respectively. Then the object spectrum can be recovered by the coherent accumulation of the object spectrum while suppressing the random multiple scattering noises in spatial frequency difference domain $\Delta \mathbf{k} = \mathbf{k}_{\text{out}} - \mathbf{k}_{\text{in}}$ as [15],
$$\tilde{O}(\Delta \mathbf{k}) \propto \sum_{\mathbf{k}_{\text{in}}} \tilde{R}(\Delta \mathbf{k} + \mathbf{k}_{\text{in}}; \mathbf{k}_{\text{in}}). \tag{2}$$
The final object image can be obtained from the inverse Fourier transform of $\tilde{O}(\Delta \mathbf{k})$. Since $\tilde{O}(\Delta \mathbf{k})$ is a function of $\Delta \mathbf{k}$, its dimension should be doubled as $2L \times 2M$ resulting in a spatial resolution up to $\lambda/4\alpha$. The image reconstruction process in Eq. (2) is illustrated in Fig. 2c.

The image reconstruction process in Eq. (2) can be efficiently calculated by transforming the output coordinate of $\widetilde{\boldsymbol{R}}$ into $\Delta \mathbf{k}$. If we define the coordinate transform $\mathcal{T}_{\Delta \mathbf{k}}$ as,
$$\mathcal{T}_{\Delta \mathbf{k}}: \quad \widetilde{\boldsymbol{R}} \to \widetilde{\boldsymbol{R}}_{\Delta \mathbf{k}} = \mathcal{T}_{\Delta \mathbf{k}}[\widetilde{\boldsymbol{R}}], \tag{3}$$
the resulting matrix $\widetilde{\boldsymbol{R}}_{\Delta \mathbf{k}}$ becomes $4LM \times LM$ as depicted in Fig. 2c. Then the summation in Eq. (2) can be calculated by adding the rows of $\widetilde{\boldsymbol{R}}_{\Delta \mathbf{k}}$ as, $\sum_{\text{row}} \widetilde{\boldsymbol{R}}_{\Delta \mathbf{k}}$. This process produces a vectorized image with $4LM \times 1$ dimension. By reshaping it into $2L \times 2M$ array, we can recover the object spectrum in the expanded bandwidth. By taking the inverse Fourier transform, we could obtain an object image with a spatial resolution of $\lambda/4\alpha$.

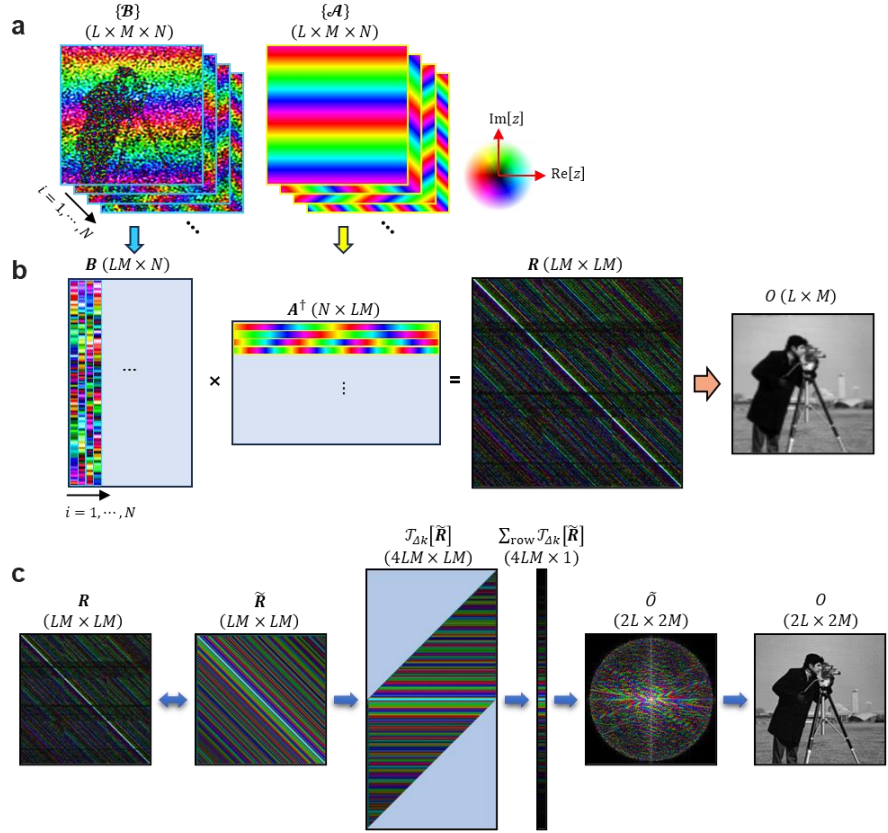

**Figure 2. Schematics of image processing in reflection matrix microscopy. a**, Illustration of the measured reflected electric field image stack $\{\mathcal{B}\}$ under a set of input field image stack $\{\mathcal{A}\}$. **b**, Construction of reflection matrix. **c**, Schematic of RMM image reconstruction in spatial frequency domain.

The image reconstruction algorithm requires summing up the shifted spectrum by $\mathcal{T}_{\Delta \mathbf{k}}$ mapping in Eq. (2). In addition, the CLASS algorithm, which will be discussed later, requires repeating this process several times during the numerical iteration process. The simplest way of this process is shifting the spectrum in each column vector of $\widetilde{R}$ by $-\mathbf{k}_{\mathrm{in}}$ as described in Fig. 3a. However, this process is time-consuming and inefficient. We found that the amount of shift in Eq. (3) is pre-determined once the matrix dimensions $L$ and $M$ are determined. Therefore, each matrix element $\widetilde{R}_{ij}$ of $\widetilde{R}$ should be mapped into a pre-determined matrix index of $\widetilde{R}_{\Delta \mathbf{k}}$. In other words, if we can pre-calibrate these indices to be mapped in $\widetilde{R}_{\Delta \mathbf{k}}$, we can perform the mapping process in Eq. (3) by simply using the logical indexing (LI) of Matlab. Figure 3b describes how we can obtain the mapping indices. As shown in Fig. 3c, this logical indexing approach is about five times faster than mapping each column one by one within a for-loop. Note that the default LI process in Matlab utilizes a single thread of CPU. This process can be even faster when we utilize multi-thread logical indexing with a MEX-file written in C language. Figure 3c compares the speed of logical indexing process of $\widetilde{R}$ with $L^2 \times L^2$ dimension for three cases: (1) mapping one-by-one with for-loop, (2) Matlab default logical indexing, and (3) multi-thread logical indexing with C-MEX file provided in Ref. [29]. The results in Fig. 3c show that the multi-thread LI can be about 25 times faster than the for-loop case when $L > 60$. Due to the finite overhead time of mex functions in Matlab (~20 ms), the default LI process in Matlab can be faster when $L < 30$. In any case, the logical indexing method was about 5 to 25 times faster than the for-loop case. However, since logical indexing requires the preparation of index values and $\widetilde{R}_{\Delta k}$, 5 times more memory than the for-loop case is necessary as presented in Fig. 3d. The detailed Matlab algorithm for the image reconstruction and logical indexing is presented in Appendix C and Ref. [29].

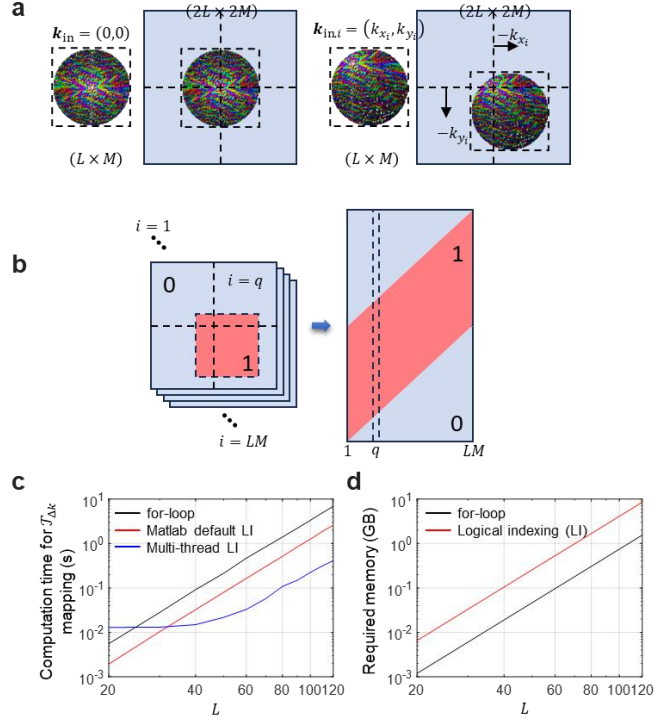

**Figure 3. Schematics of logical indexing for Δk coordinate mapping. a**, Examples of **Δk** space mapping by shifting the electric field spectrum with $-\boldsymbol{k}_{\text{in},i}$. **b**, Schematic of finding index values of **Δk** mapping. **c-d**, Performance of logical indexing in comparison with utilizing for-loop.

## 4. 3D reconstruction from reflection matrix

Since reflection matrix microscopy measures both the amplitude and phase of the reflected electric field from the sample, it provides volumetric information of the target. The free space propagation of the electric field can be calculated using the angular spectrum method as [31],

$$\tilde{E}(k_x, k_y; d) = \tilde{E}(k_x, k_y; 0) \cdot \exp\left[id\sqrt{n_0^2 k_0^2 - k_x^2 - k_y^2}\right] \quad (4)$$

where $\tilde{E}(k_x, k_y; d)$ is the electric field spectrum at $z = d$ plane, $k_0$ is the vacuum wavenumber, and $n_0$ is the refractive index of the surrounding medium.

Using Eq. (4), we can transform the reflection matrix $\boldsymbol{R}$ measured at the $z = 0$ plane into $\boldsymbol{R}_d$ at the $z = d$ plane. As illustrated in Fig. 4a, if we assume a point particle located at the $z = d$ plane, the incident field hitting the particle can be described by propagating the incident field forward with a distance $d$. Meanwhile, the spherical wave reflected from the particle is focused on a position $d$ in front of the camera. In other words, the image of the particle will be defocused by an amount of $d$. Therefore, we need to back-propagate the measured electric field by a distance $d$.

As a result, as discussed in Appendix B, the transformed reflection matrix $\widetilde{\boldsymbol{R}}_d$ in the spatial frequency domain can be written as

$$\widetilde{\boldsymbol{R}}_d = \boldsymbol{H}_{-d}\widetilde{\boldsymbol{R}}\boldsymbol{H}_{-d} \tag{5}$$

where $\boldsymbol{H}_{-d}$ is a diagonal matrix with diagonal elements given by $\exp\left(-id\sqrt{n_0^2 k_0^2 - k_x^2 - k_y^2}\right)$. Figure 4c presents the map of $\boldsymbol{H}_{-d}$ with $d = 1$ μm. By substituting $\widetilde{\boldsymbol{R}}_d$ into Eq. (2), we can reconstruct the RMM image at $z = d$ plane. For a volumetric sample, we can scan $d$ to obtain 3D RMM images. Note that if a broadband light source is used for temporal coherence gating, the 3D reconstruction of the target object is only possible within the coherence length of the light source.

We present an example of the 3D reconstruction in Fig. 4d-j. First, we numerically simulate a reflection matrix of point particles distributed in a 3D volume of 25×25×12 μm³ with an illumination wavelength of 1 μm and a numerical aperture $\alpha = 1.2$. Using Eqs. (2) and (4), we reconstructed RMM images at different depths, as depicted in Fig. 4d. Additionally, we determined the spatial resolution of the 3D reconstruction by calculating the full width at half maximum (FWHM) of a particle located within the white dashed box in Fig. 4d. As shown in Fig. 4g-h, the resulting FWHM was 306 nm in the lateral direction and 845 nm along the axial direction.

To verify the accuracy of the 3D reconstruction, we compared the maximum intensity projection (MIP) image of the reconstructed 3D RMM images with the ground-truth target used for the simulation, as presented in Fig. 4i-j with color-coded depth. Although individual particles are blurred due to the finite numerical aperture of the system, the locations of individual particles perfectly match the ground-truth, implying the accuracy of the reconstruction. The numerically generated reflection matrix data and 3D reconstruction code used in Figure 4 are provided in the software package in Ref. [29].

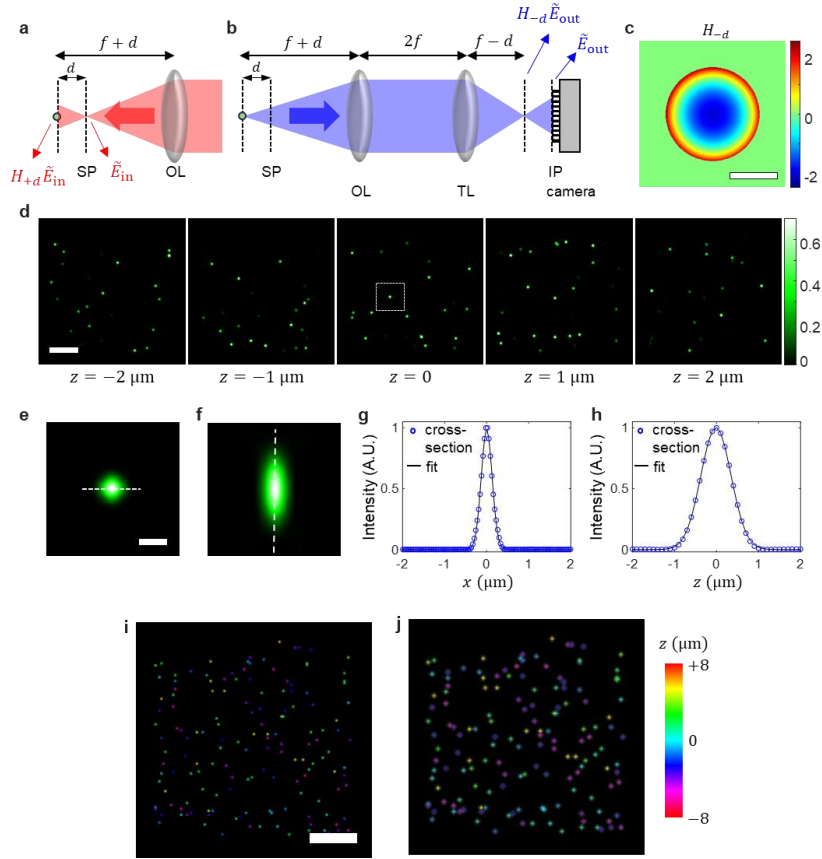

**Figure 4. 3D reconstruction in reflection matrix microscopy. a-b,** Schematic of input and output electric field propagation for detecting a particle located in $z = d$ plane. **c,** Transfer function $H_{-d}$ of angular spectrum method when $d = 1$ µm. Scale bar: $k_0\alpha$. **d,** 3D reconstruction result with a numerically simulated reflection matrix of point particles dispersed in 3D space. Scale bar: 5 µm. **e-f,** x-y and x-z cross-sectional images of a point particle depicted by a white dashed box in **d**. Scale bar: 500 nm. **g-h,** Cross-section of dashed line in **e-f**. **i-j,** Maximum intensity projection image of the ground-truth target, and reflection matrix-based 3D reconstruction, respectively, with color-coded depth. Scale bar: 5 µm.

## 5. Aberration correction by CLASS algorithm

In optical imaging, the sample-induced aberrations cause the distortion of the image and the degradation of the contrast. In general, such aberrations can be described as the angle-dependent phase shifts during the illumination process ($\theta_{in}(\mathbf{k}_{in})$), and reflection process ($\theta_{out}(\mathbf{k}_{out})$). Under these input and output aberrations, the reflection matrix elements in Eq. (1) are modified as [16],

$$\tilde{R}(\mathbf{k}_{out}; \mathbf{k}_{in}) = \gamma P_{out}(\mathbf{k}_{out})\tilde{O}(\mathbf{k}_{out} - \mathbf{k}_{in})P_{in}(\mathbf{k}_{in}) + \tilde{M}(\mathbf{k}_{out}; \mathbf{k}_{in}). \quad (6)$$

Here, $P_{in}(\mathbf{k}_{in}) = \text{circ}(|\mathbf{k}_{in}|/k_0\alpha)\exp[i\theta_{in}(\mathbf{k}_{in})]$ and $P_{out}(\mathbf{k}_{out}) = \text{circ}(|\mathbf{k}_{out}|/k_0\alpha) \cdot \exp[i\theta_{out}(\mathbf{k}_{out})]$ are input and output complex pupil functions, respectively, where $\text{circ}(r)$ is circle function which has unit value for $0 \leq r \leq 1$ and 0 otherwise. In matrix expression, Eq. (6) can be written as $\tilde{\mathbf{R}} = \mathbf{P}_{out}\tilde{\mathbf{O}}\mathbf{P}_{in} + \tilde{\mathbf{M}}$, where $\mathbf{P}_{in}$ and $\mathbf{P}_{out}$ are diagonal matrices whose diagonal elements are given by $P_{in}(\mathbf{k}_{in})$ and $P_{out}(\mathbf{k}_{out})$, respectively, and $\tilde{\mathbf{M}}$ is the multiple scattering matrix made of $\tilde{M}(\mathbf{k}_{out}; \mathbf{k}_{in})$.

The CLASS algorithm can computationally find the input and output aberrations from a reflection matrix with an iteration process. The overall flow-chart of the algorithm is presented in Fig. 5a. First, if we transform the reflection matrix in Eq. (6) into $\Delta \mathbf{k}$ space as defined in Eq. (3), the matrix elements of $\widetilde{\mathbf{R}}_{\Delta \mathbf{k}}$ can be written as,

$$\widetilde{R}_{\Delta \mathbf{k}}(\Delta \mathbf{k}; \mathbf{k}_{in}) = \gamma P_{out}(\Delta \mathbf{k} + \mathbf{k}_{in})\widetilde{O}(\Delta \mathbf{k})P_{in}(\mathbf{k}_{in}) + \widetilde{M}(\Delta \mathbf{k}; \mathbf{k}_{in}) \qquad (7)$$

Regardless of the column index $i$, columns of $\widetilde{\mathbf{R}}_{\Delta \mathbf{k}}$ is proportional to $\widetilde{O}(\Delta \mathbf{k})$ with a scalar phase shift value if we ignore the multiple scattering components. This means finite correlations exist between individual columns of $\widetilde{\mathbf{R}}_{\Delta \mathbf{k}}$. In fact, the angle of the correlation $\theta_{in}^{(1)}(\mathbf{k}_{in})$ among the columns corresponds to the approximate estimate of $\theta_{in}(\mathbf{k}_{in})$. The reason $\theta_{in}^{(1)}(\mathbf{k}_{in})$ is not equal to $\theta_{in}(\mathbf{k}_{in})$ is due to the presence of output aberration and multiple scattering. Here, we use a superscript (1) to represent the iteration number. As illustrated in the first row of Fig. 5a, once we find the phase values $\theta_{in}^{(1)}$, we can correct the phase by multiplying a diagonal matrix $\mathbf{P}_{in}^{(1)}$, whose diagonal elements are $e^{-i\theta_{in}^{(1)}(\mathbf{k}_{in})}$, with $\widetilde{\mathbf{R}}$. Similarly, we can correct the output phase shift $\theta_{out}$ by repeating the same process with a transposed reflection matrix $\widetilde{\mathbf{R}}^T$. The first-round estimation of the phase shift maps, $\theta_{in}^{(1)}$ and $\theta_{out}^{(1)}$, are not accurate. In Eq. (7), while quantifying $\theta_{in}^{(1)}$ by comparing the columns of $\widetilde{\mathbf{R}}_{\Delta \mathbf{k}}$, the presence of output aberration $\theta_{out}$ will cause errors for finding $\theta_{in}^{(1)}$. Similarly, while calculating $\theta_{out}^{(1)}$, the residual input aberration $\theta_{in} - \theta_{in}^{(1)}$ after the input correction process will cause errors. Therefore, we need to iterate the whole process until $\theta_{in}^{(j)}$ and $\theta_{out}^{(j)}$ for $j^{th}$ iteration step converges to zero. The whole iteration process is indicated in the black arrows of Fig. 5a.

After finishing the CLASS iterations, we can find the input and output aberration by summing up the phase maps quantified in all the iterations steps, i.e. $\theta_{in}^{CLASS} = \sum_j \theta_{in}^{(j)}$ and $\theta_{out}^{CLASS} = \sum_j \theta_{out}^{(j)}$. Also, we can reconstruct the CLASS image from Eq. (2) with aberration-corrected reflection matrix $\widetilde{\mathbf{R}}^{CLASS} = \mathbf{P}_{out}^{CLASS}\widetilde{\mathbf{R}}\mathbf{P}_{in}^{CLASS}$ with correction matrices $\mathbf{P}_{in}^{CLASS} = \exp[-i\theta_{in}^{CLASS}]$ and $\mathbf{P}_{out}^{CLASS} = \exp[-i\theta_{out}^{CLASS}]$. Figure 5b shows an example of the CLASS algorithm with a numerical simulation data provided in the software package in Ref. [29].

In the original CLASS paper of Ref. [16], we described the phase quantification process of $\theta_{in}^{(j)}$ and $\theta_{out}^{(j)}$ by comparing the columns of $\widetilde{\mathbf{R}}_{\Delta \mathbf{k}}$ one by one. However, this is time-consuming and less accurate, especially in the presence of multiple scattering. We streamlined our algorithm to find the estimates of $\theta_{in}^{(j)}$ for all the columns in $\widetilde{\mathbf{R}}_{\Delta \mathbf{k}}$ at once by solving

$$\arg\min_{\theta^{(j)}} \left\| \widetilde{\mathbf{R}}_{\Delta \mathbf{k}} - \tau \widetilde{\mathbf{o}}^T \times e^{i\theta_{in}^{(j)}} \right\|_2, \qquad (8)$$

where $\widetilde{\mathbf{o}}^T(\Delta \mathbf{k})$ is a transpose of the vectorized object spectrum in $\Delta \mathbf{k}$ space (column vector), $e^{i\theta_{in}^{(j)}}$ is vectorized phase errors (row vector), and $\|A\|_2$ implies the Frobenius norm of a matrix $A$. We solved this minimization problem with a well-known power iteration method [32] which is composed of simple matrix–vector multiplications as described in Table 1.

> **Input**: matrix $\widetilde{R}_{\Delta k}$ in $\Delta k$ space, the iteration number $n_p$.
> 1. Let $v_0 = 1$ as an initial column vector.
> 2. For $l = 1$ to $n_p$
>    a. $v_l = (\widetilde{R}_{\Delta k})^\dagger (\widetilde{R}_{\Delta k} v_{l-1})$ where $(\widetilde{R}_{\Delta k})^\dagger$ is the conjugate transpose of $\widetilde{R}_{\Delta k}$.
>    b. $v_l = v_l / |v_l|$: normalize to unit amplitude.
> 
> **Output**: phase $\theta_{\text{in}}^{(j)} = \arg(v_{n_p})$.

**Table 1. Power iteration algorithm**

It is noteworthy that the power iteration method described in the above algorithm is known to be identical to finding the first singular vector of $\widetilde{R}_{\Delta k}$ in singular value decomposition (SVD). There have been developed many advanced algorithms for SVD, and the power iteration method is regarded as a less efficient algorithm than SVD. However, in the CLASS algorithm, only the first singular vector is of interest, and the power iteration method is better suited.

The convergence speed of the power iteration method is related to the ratio between the first singular value of $\widetilde{R}_{\Delta k}$ and other values. Therefore, in case of strong multiple scattering, which contributes to the other singular vectors, the convergence speed is slower, which necessitates the increase of the iteration number $n_p$. Also, the CLASS algorithm converges only when the first singular vector $\widetilde{R}_{\Delta k}$ contains the object information. We present the Matlab algorithm for the CLASS and power iteration in Algorithm 3-4 of Appendix C.

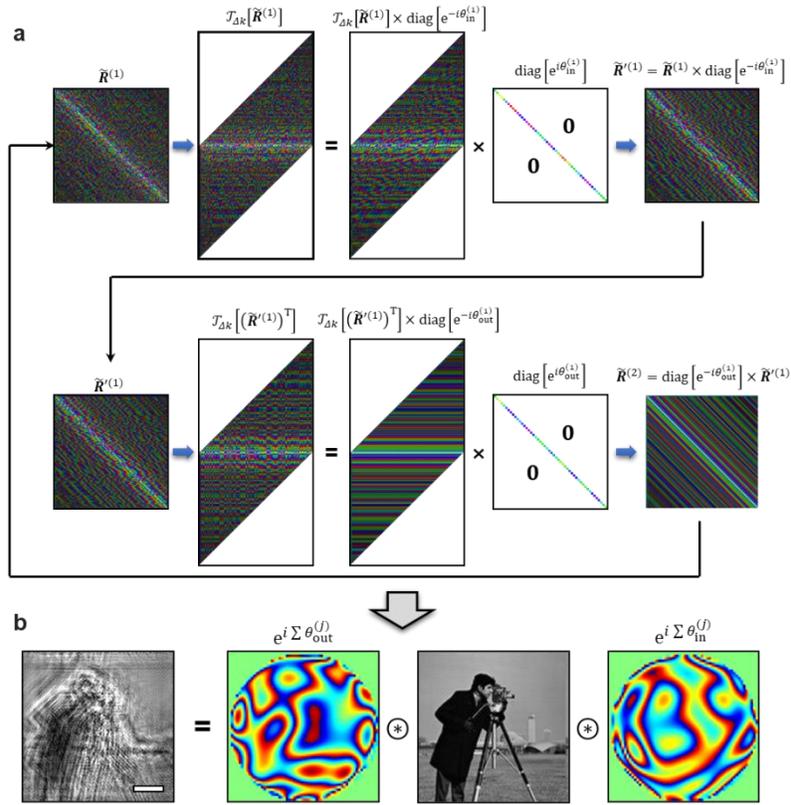

**Figure 5. Schematic of CLASS algorithm. a**, Flow-chart of CLASS algorithm. Initial reflection matrix $\widetilde{R}^{(1)}$ is mapped into $\Delta \mathbf{k}$ space, and then decomposed into a diagonal phase modulation matrix $P_{in}^{(1)}$ and the residual matrix. After correcting the input aberration by multiplying $P_{in}^{(1)}$ in the right-hand side of $\widetilde{R}^{(1)}$, repeat the same procedure for finding $P_{out}^{(1)}$ with transposed matrices. Iterate the whole process until the updated phase maps $\theta_{in}^{(N)}$ and $\theta_{out}^{(N)}$ converge. **b**. An example of the CLASS algorithm with simulation data provided in Ref. [29].

## 6. Implementation of the CLASS algorithm with experimental data

In Fig. 6, we present an example of the CLASS algorithm application with real experimental data. As an imaging target, a USAF 1951 resolution target was placed under 1.5 mm thick glass plate. Before the aberration correction, the initial RMM image of the target was highly distorted due to the strong spherical aberration induced by the thick glass plate (Fig. 6b). By applying the CLASS algorithm, the input and output aberrations were quantified within 10 iteration steps as displayed in Fig. 6a. Figures 6c-d show the resulting input and output aberration maps and CLASS image, respectively. Note that the input aberration map has fast varying fluctuations over the circularly symmetric spherical aberration due to the drift of reference pathlength during the reflection matrix measurement. Since the matrix was measured in plane wave basis, such phase drift was super-imposed in the input aberration map. We provide the experimental data in Fig. 6 as a test data in the software package of Ref. [29].

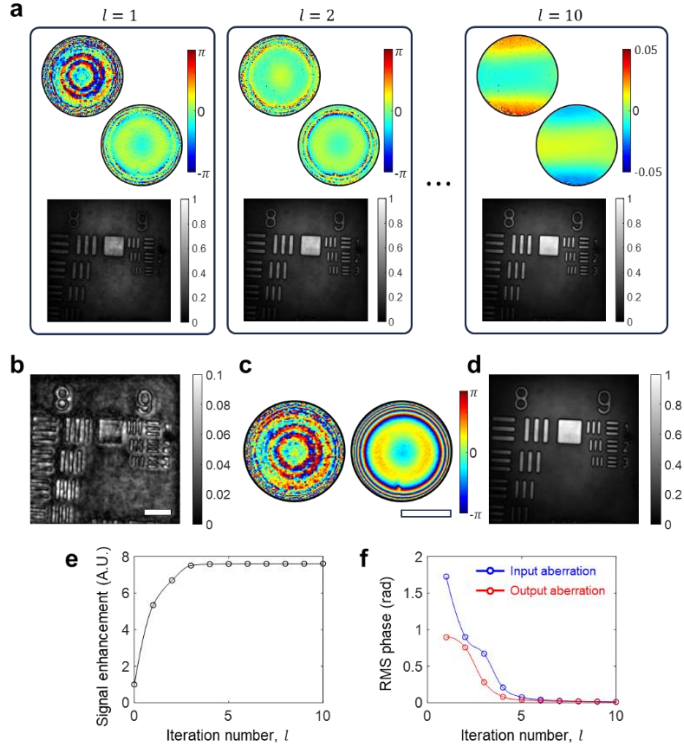

**Figure 6. Implementation of CLASS algorithm. a**, Input/output aberration maps and the CLASS image for the representative iteration steps of $l = 1, 2,$ and $10$. **b**, RMM image of the target before applying the CLASS algorithm. Scale bar: 5 µm. **c**, Input and output aberration maps quantified by the CLASS algorithm. Scale bar: $k_0\alpha$ with $\alpha = 0.8$, $k_0 = 2\pi/0.8$ rad/µm. **d**, CLASS image after aberration correction. **e**, Total intensity enhancement with iterations. **f**, RMS phase values of input and output aberration maps updated with iterations.

## 7. Low spatial frequency blocking for faster CLASS convergence

The CLASS algorithm can distinguish the input and output aberrations independently. However, when low spatial frequency components of the target object spectrum are dominant, the convergence of the CLASS algorithm is slow and often fails to work. To illustrate this, let us decompose the object matrix $\widetilde{O}$ into an identity matrix and a higher frequency matrix $\widetilde{O}^{\text{HF}}$, i.e. $\widetilde{O} = I + \widetilde{O}^{\text{HF}}$. Then, the reflection matrix is written as, $\widetilde{R} = P_{\text{out}}P_{\text{in}} + P_{\text{out}}\widetilde{O}^{\text{HF}}P_{\text{in}} + M$. If the contribution of high spatial frequency in $\widetilde{O}^{\text{HF}}$ is negligible, the CLASS algorithm will find $P_{\text{out}}P_{\text{in}}$ as an input aberration in the first iteration step and will stop iterates. As an extreme case, if the target is an ideal mirror, i.e. $\widetilde{O} = I$, and the CLASS algorithm fails to distinguish the input and output aberrations. To avoid such problems, blocking the low spatial frequency of $\widetilde{R}_{\Delta\mathbf{k}}$ in the $\Delta k$ space by multiplying a mask function $H(\Delta\mathbf{k}) = \begin{cases} 0, |\Delta\mathbf{k}| \leq \epsilon \\ 1, |\Delta\mathbf{k}| > \epsilon \end{cases}$ is necessary while calculating the power iteration in Eq. (8).

We present the performance of the low spatial frequency blocking in Fig. 7. Without blocking, CLASS algorithm does not work properly. On the other hand, by increasing the block size, the convergence speed of the CLASS algorithm was greatly increased. Note that the low-spatial frequency blocking is applied only at the step of finding the aberrations, and full spectra are used at the time of image reconstruction.

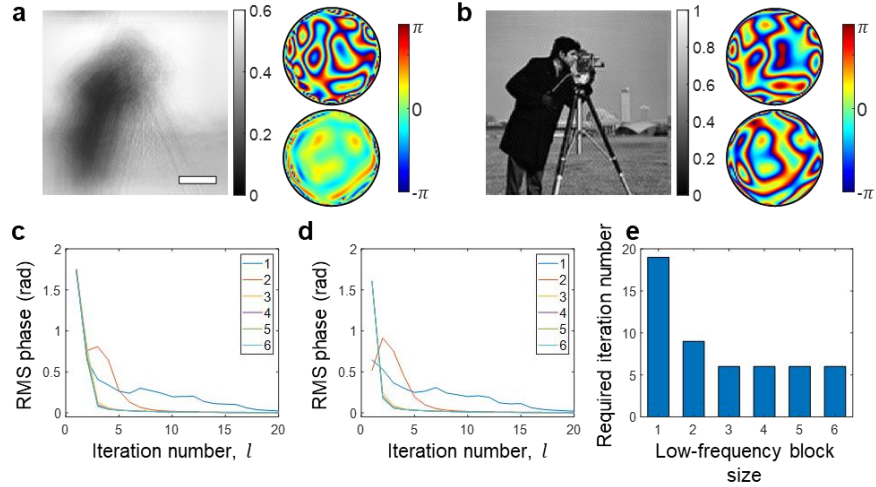

**Figure 7. The effect of low spatial frequency blocking. a**, CLASS image, and input/output aberration maps without low-spatial frequency blocking. **b**, Same as (**a**), but after low-spatial frequency blocking. **c-d**, RMS phase of input and output aberration maps while iteration, respectively, with different low-spatial frequency block sizes $\epsilon$, whose unit is given by $2\pi/L$ with detection area $L$. **e**, The number of required iterations for the RMS phase less than 0.03 rad at different low-spatial frequency block sizes.

## 8. Discussions

Reflection matrix microscopy and its associated algorithms have provided a powerful imaging platform capable of computationally correcting complex aberrations even in strong multiple scattering noise. This platform has been further advanced in recent years to trace and correct multiple scattering, enabling *in vivo* through-skull imaging up to cortical layer 4 without losing resolving power and extending imaging depths for super-resolution imaging [21, 23]. Despite its advantages, this microscopy method is relatively complex to adopt compared to conventional microscopy. The primary challenge lies in the higher data dimensions of the reflection matrix and the consequent conceptual and technical difficulties in implementing the image reconstruction algorithms.

    This study introduces core matrix processing engines for handling reflection matrices in a more comprehensible manner and presents significant algorithmic advancements that substantially increase processing speed. The proposed techniques, such as logical indexing, power iterations, and low-spatial frequency blocking, have improved image reconstruction speed by tens to hundreds of times compared to the original algorithm first introduced in Ref. [16]. These advancements are expected to facilitate real-time applications in dynamic biological studies and expedite the use of RMM as a wavefront sensor for hardware AO in multiphoton imaging and SMLM. To enable researchers to effectively utilize these advanced technologies, we have provided ready-to-use Matlab implementations, along with simulation and experimental data [29].

    The algorithms presented in this study have served as fundamental engines for more advanced algorithms such as conjugate CLASS [21], flexible holographic endoscope

reconstruction algorithms [33], and multiple scattering tracing algorithms [22], providing significant benefits to researchers utilizing these advanced techniques. Future algorithmic improvements are still necessary; for example, these algorithms have high memory requirements, necessitating the development of effective strategies for processing larger areas simultaneously [27]. Implementing logical indexing in CUDA could further enhance speed, ultimately achieving near real-time correction, which would be highly beneficial for *in vivo* studies or hardware AO applications.

## 9. Appendix

### A. Basis transform of reflection matrix

With a reflection matrix $\boldsymbol{R}$, the reflected electric field vector $\boldsymbol{u}$ from a sample under incident electric field vector $\boldsymbol{v}$ can be written as,

$$\boldsymbol{u} = \boldsymbol{R}\boldsymbol{v}. \tag{9}$$

If the reflection matrix is constructed under a position basis, the $i^{\text{th}}$ elements of $\boldsymbol{u}$ and $\boldsymbol{v}$ correspond to the electric field components at that position coordinate. By the linear superposition principle of electromagnetic waves, the input and output bases of the reflection matrix can be transformed into different bases with unitary transformations. Considering the transformation of the input and output bases defined by the transform matrices $\boldsymbol{H}_{\text{in}}$ and $\boldsymbol{H}_{\text{out}}$, the input and output basis of the transformed coordinate $(\boldsymbol{u}', \boldsymbol{v}')$ can be obtained by $\boldsymbol{u}' = \boldsymbol{H}_{\text{out}}\boldsymbol{u}$ and $\boldsymbol{v}' = \boldsymbol{H}_{\text{in}}\boldsymbol{v}$, respectively. Therefore, the reflection matrix in the transformed basis can be described by,

$$\boldsymbol{u}' = \boldsymbol{H}_{\text{out}}\boldsymbol{u} = \boldsymbol{H}_{\text{out}}\boldsymbol{R}\boldsymbol{v} = (\boldsymbol{H}_{\text{out}}\boldsymbol{R}\boldsymbol{H}_{\text{in}}^{-1})\boldsymbol{v}' = \boldsymbol{R}'\boldsymbol{v}' \tag{10}$$

If $\boldsymbol{H}_{\text{in}}$ is a unitary matrix $\boldsymbol{H}_{\text{in}}^{-1} = \boldsymbol{H}_{\text{in}}^{\dagger}$. For example, if $\boldsymbol{H}_{\text{in}}$ and $\boldsymbol{H}_{\text{out}}$ are discrete Fourier transform matrices $\boldsymbol{D}$, then the reflection matrix $\widetilde{\boldsymbol{R}}$ in the angular spectrum basis can be written as, $\widetilde{\boldsymbol{R}} = \boldsymbol{D}\boldsymbol{R}\boldsymbol{D}^{\dagger}$. As discussed in the main text, in general, fast Fourier transformation (FFT) algorithm is much faster than the matrix multiplication approach. Therefore, we applied the FFT algorithm instead of multiplying $\boldsymbol{D}$ for calculating $\widetilde{\boldsymbol{R}}$ from $\boldsymbol{R}$. The Matlab function of this process is given in the software packages of Ref. [29].

### B. 3D propagation of the reflection matrix

The free space propagation of electric field can be calculated by Rayleigh – Sommerfeld diffraction theory [31] as,

$$E(x, y, d) = \frac{1}{i\lambda} \iint_{-\infty}^{\infty} E(x', y', 0) \frac{e^{in_0 k_0 r} d}{r^2} dx' dy' \tag{11}$$

where $E(x', y', 0)$ is electric field at $z = 0$ plane, $d$ is the propagation distance, $n_0$ is the refractive index of the medium, $k_0 = 2\pi/\lambda$ is vacuum wavenumber, and $r = \sqrt{(x-x')^2 + (y-y')^2 + d^2}$. Eq. (11) can be efficiently calculated by the convolution theorem as,

$$\widetilde{E}(k_x, k_y; d) = \widetilde{E}(k_x, k_y; 0) \cdot \exp\left[id\sqrt{n_0^2 k_0^2 - k_x^2 - k_y^2}\right] \tag{12}$$

where $\widetilde{E}$ is 2D Fourier transform of $E$ along $(x, y)$ coordinate, and $(k_x, k_y)$ is lateral wavevector.

Equation (12) can be written in a vectorized form with a diagonal transfer matrix $\boldsymbol{H}_d$ as,

$$\widetilde{\boldsymbol{v}}_d = \boldsymbol{H}_d \widetilde{\boldsymbol{v}} \tag{13}$$

where the diagonal elements of $H_d$ is given by $\exp\left(id\sqrt{n_0^2 k_0^2 - k_x^2 - k_y^2}\right)$. Since $H_d$ is a unitary matrix, the free-space propagation of the electric field can be considered as a unitary transformation of $\widetilde{v}$ into $\widetilde{v}_d$.

If a reflection matrix $R$ is measured at a specific focal plane at $z = 0$. Then the reflection matrix at $z = d$ plane can be constructed from $R$ by applying free-space propagation discussed in Eqs. (11-13). At the propagated plane, the input basis of the initial reflection matrix $R$ should be propagated by an amount of $d$. In the meanwhile, the reflection signal originating from the $z = d$ plane is de-focused at the camera plane. Therefore, the output basis of $R$ should be back-propagated by $d$. In other words, the input basis should be transformed with unitary matrix $H_d$ while the output basis should be transformed with $H_{-d}$. Then the reflection matrix at $z = d$ plane can be written as $\widetilde{u}_d = H_{-d}\widetilde{R}H_{-d}\widetilde{v}_d$ by Eq. (10). Finally, the reflection matrix at propagated plane can be calculated by multiplying a diagonal matrix $H_{-d}$ in both sides of $\widetilde{R}$. Then the target image at distance $d$ from the initial focus plane can be reconstructed from the propagated matrix with Eqs. (2-3) in the main text.

*C. Matlab algorithms for reflection matrix microscopy*

In this section, we summarize Matlab algorithms of reflection matrix microscopy.

**Algorithm 1. Image reconstruction from $\widetilde{R}$**

```
1.   function image = image_reconstruction(Rk, Nx, Ny)
2.       indx_dk = get_dk_logical_index(Nx, Ny);
3.       dk_matrix = zeros (Nx*Ny*4, Nx*Ny, 'single');
4.       % default Matlab logical indexing
5.       dk_matrix(indx_dk) = Rk;
6.       image_k=(sum(dk_matrix, 2));
7.       image_k = reshape(image_k, Ny*2, Nx*2);
8.       img=ifftshift(ifft2(ifftshift(imgk)));
9.   end
```

**Algorithm 2. Calculate indices of $\mathcal{T}_{\Delta k}$ mapping**

```
1.   function indx_dk = get_dk_logical_index(Nx, Ny)
2.       Nky=Ny/2;
3.       Nkx=Nx/2;
4.       [kx, ky]=meshgrid(-Nkx:Nkx-1, -Nky:Nky-1);
5.       kx=kx(:);
6.       ky=ky(:);
7.       dk_matrix=zeros(16*Nkx*Nky, 4*Nkx*Nky);
8.       kmap=padarray(ones(Nky*2, Nkx*2), [Nky, Nkx]);
9.       for ii=1:4*Nkx*Nky
10.          shifted_kmap=circshift(kmap, [-ky(ii), -kx(ii)]);
```

11.         dk_matrix(:, ii)=shifted_kmap(:);
12.     end
13.     indx_dk = uint32(find(dk_matrix==1));
14. end

**Algorithm 3. Power iteration**

1.    function v = Power_Iteration(T, max_PI_num)
2.        m = size(T, 2);   % number of input basis
3.        v = ones(m, 1)/sqrt(m);   % initial phase vector
4.        for ii = 1:max_PI_num
5.          v = T'*(T*v);
6.          v = v./abs(v);   % for finding phase only solution
7.        end
8.    end

**Algorithm 4. CLASS algorithm**

1.    function [CLASS_image, ab_in, ab_out] = CLASS_function(Rk, Nx, Ny)
2.        ab_in = ones(Ny, Nx);
3.        ab_out = ones(Ny, Nx);
4.        indx_dk = get_dk_logical_index(Nx, Ny);
5.
6.        for ii = 1:max_iteration_number
7.         % find input correction
8.         dk_matrix = zeros (Nx*Ny*4, Nx*Ny, 'single');
9.         dk_matrix(indx_dk) = Rk;
10.       input_phase = Power_Iteration(dk_matrix, max_PI_num);
11.       ab_in = ab_in .* reshape(input_phase, Ny, Nx);
12.
13.       % correct input aberration
14.       Rk = Rk.*input_phase(:).';
15.
16.       % find output aberration
17.       % (same process with input case with transposed reflection matrix)
18.       dk_matrix = zeros (Nx*Ny*4, Nx*Ny, 'single');
19.       dk_matrix(indx_dk) = Rk.';   % transpose of reflection
20.       output_phase = Power_Iteration(dk_matrix, max_PI_num);
21.       ab_out = ab_out .* reshape(output_phase, Ny, Nx);
22.
23.       % correct output aberration

```
24.         Rk = output_phase(:).*Rk;
25.     end
26.
27.     % generate final CLASS image using Algorithm 1
28.     CLASS_image = image_reconstruction(Rk, Nx, Ny)
29. end
```

## Acknowledgments

This work was supported by Institute for Basic Science (IBS-R023-D1) and National Research Foundation of Korea (NRF) Grant Funded by the Korean Government (MSIT) (RS-2023-00213310).

## Disclosures

The authors declare no conflicts of interest.

## Data availability

Data and Matlab codes underlying the results presented in this paper are available in Ref. [29].